# Resource Matchmaking Algorithm using Dynamic Rough Set in Grid Environment

Iraj Ataollahi
Computer Engineering Department
Iran University of Science and Technology
Tehran, Iran
ir_ataollahi@mail.iust.ac.ir

Morteza Analoui
Computer Engineering Department
Iran University of Science and Technology
Tehran, Iran
analoui@iust.ac.ir

*Abstract*— Grid environment is a service oriented infrastructure in which many heterogeneous resources participate to provide the high performance computation. One of the bug issues in the grid environment is the vagueness and uncertainty between advertised resources and requested resources. Furthermore, in an environment such as grid dynamicity is considered as a crucial issue which must be dealt with. Classical rough set have been used to deal with the uncertainty and vagueness. But it can just be used on the static systems and can not support dynamicity in a system. In this work we propose a solution, called Dynamic Rough Set Resource Discovery (DRSRD), for dealing with cases of vagueness and uncertainty problems based on Dynamic rough set theory which considers dynamic features in this environment. In this way, requested resource properties have a weight as priority according to which resource matchmaking and ranking process is done. We also report the result of the solution obtained from the simulation in GridSim simulator. The comparison has been made between DRSRD, classical rough set theory based algorithm, and UDDI and OWL-S combined algorithm. DRSRD shows much better precision for the cases with vagueness and uncertainty in a dynamic system such as the grid rather than the classical rough set theory based algorithm, and UDDI and OWL-S combined algorithm.

## I. INTRODUCTION (HEADING 1)

Nowadays, Grid is considered as a service-oriented computing infrastructure [1]. Open Grid Services Architecture (OGSA) [2], which has been promoted by Global Grid Forum, has been used for dealing with service-oriented problem [3].
Many resources such as workstations, clusters, and mainframes with various properties such as main memory, CPU speed, bandwidth, virtual memory, hard disk, operating system, CPU vender, number of CPU elements etc are joining and leaving the grid environment. On the Other hand many users want to use these resources to run their jobs with different requirements. But there are always differences between which a user requested and whitch have been registered in a grid GIS. To solve this vagueness and uncertainty we use rough set theory, proposed by Z. Pawlak in 1982 [4], which has been used in vast area of computer science such as data mining, pattern recognition, machine learning and knowledge acquisition etc [5].

One of the first methods that can be used for service discovery is UDDI which is used for web service publication and discovery. The current web service discovery mechanism is based on the standard of UDDI [6]. In UDDI, XML is used to describe data in business services. UDDI process Searchs queries according to keywords and classification information. There is limitation with the discovery mechanism of UDDI. Firstly, machine can read XML data, but it can not understand XML data. Different query keywords may be semantically equivalent, whereas UDDI can not infer any information from keywords or tModels it can easily make mistake. Secondly, search by keywords and taxonomy is not suitable for web service discovery. Furthermore, UDDI does not support search by service capabilities and other properties [7]. This makes UDDI search method a low precision method [6].

By advent of semantic web, services can be annotated with metadata for enhancement of service discovery. One of the earliest to add semantic information is DAML-S [8]. DAML-S uses semantic information for discovering Web services. DAML-S uses ontological description to express web service capacity and character.

OWL-S is an OWL [9] based ontology for encoding properties of Web services. OWL-S technology is used to facilitate service annotation and matching. OWL-S ontology defines a service profile for encoding a service description, a service model for specifying the behavior of a service, and service grounding for how to invoke the service. Actually, by using domain ontology descried in OWL, using special software such as protégé [10], a service discovery process involves a matching between the profile of a service advertisement and the profile of a service request. The service profile describes the functional properties such as inputs, outputs, preconditions, and effects, and non functional properties such as service name, service category, and aspects related to the quality of service.

In [11] a quantification standard for semantic service matching has been presented that modifies the classical matching algorithm based on OWL-S. Matching algorithm has used the quantification standard of service matching and OWL-WS. In [12] service composition algorithm has constructed a mathematical model and converted it to the shortest path problem in order to find process that can satisfy customer need in best conditions.

In [7] an approach has been developed for integrating semantic features into UDDI. The approach uses a semantic matchmaker that imports OWL-S based semantic markups and service properties into UDDI. The combination of OWL-S and UDDI shows there could be a service discovery which supports web service expression while UDDI is used. The matchmaker, therefore, enables UDDI to store semantic information of web services and process service search queries based on semantic similarity of web service properties [7].

The above-mentioned methods facilitate service discovery in some way. However, when matching service advertisements with service requests, these methods assume that service advertisements and service requests use consistent properties to describe relevant services. But for a system such as Grid with a large number of resources and users which have their own predefined properties to describe services, it can't be true that service advertisements and service requests use consistent properties to describe services. In other words, some properties may be used in service advertisement that may not be used by service request. So, an approach must be taken into consideration to deal with uncertainty of service properties when matching service advertisements with service requests.

Rough set theory is a new mathematical theory which deals with uncertainty and vagueness [13]. In addition to the use of rough set theory, we use service ontology to describe resources in a classified form. This ontology has been made according to the Karlsruhe ontology model [10].

The remainder of this paper is organized as fallows. Part II is a description of rough set theory, part II is a description of the algorithm implemented and used in this paper, part IV is a comparison of our algorithm with UDDI and OWL-S combined model proposed in [14] and rough set based matchmaking algorithm [18], and finally part V is the conclusion and future works.

## II. RELATED WORKS

While the grid environment moves towards a service-oriented computing infrastructure, service discovery is becoming a vital part of this environment. One of the earliest methods for service publication and discovery is UDDI which only supports keyword matches and does not support any semantic service. DAML-S is the earliest to add semantic information for discovering web services [15]. DAML-S offers enough semantic information expressing Web service capacity and character with ontological description of web services. In past few years, a great amount of studies have been carried out on the basis of OWL-S, such as semantic expression service bundling [16], ontology-based service matching [16], OWL-S and UDDI combination [14]. In the [17] a metric is proposed to measure the similarity of semantic services annotated with OWL ontology. Similarity is calculated by defining the intrinsic information value of a service description based on the inferencibility of each of OWL constructs. All the above methods do not support uncertainty in properties. Rough set theory is used for dealing with vagueness and missing data in large variety of domains. So, compared with the work mentioned above, rough set theory can tolerate uncertain properties in matching resources. In [18] we have proposed a rough set based algorithm to deal with uncertainty and vagueness. In this paper, our algorithm works in two steps. The First step is dependent properties reduction which removes dependent properties. The Second step is matchmaking which matches and ranks resources according to requested resource.

## III. CLASSICAL ROUGH SET THEORY

Rough set theory which is proposed by Pawlak, in 1982, has been proved to be a good mathematical tool to describe and model uncertainty and imprecision. It has been widely applied in artificial intelligent, pattern recognition, data mining, fault diagnostics etc [19]. There are many advantages of rough sets theory; for example, no preliminary or additional information is needed and only the facts in the data are considered.

Fig. 1 [18] shows that rough set is based on the concept of an upper and a lower approximation of a set. For a given set X the yellow grids represent its upper approximation of set X, and the green grids represent the lower approximation of set X and the black line represents the boundary region of set X.

Let:

- U: a set of N registered resources, $U = \{u_1, u_2, \ldots, u_N\}$, $N \geq 1$.
- P: a set of M properties used to describe the N registered resources of the set U, $P = \{p_1, p_2, \ldots, p_M\}$, $M \geq 2$.
- Q: a set of M registered resource properties relevant to a resource request R in terms of resource ontology whose irrelevant properties have been removed, $Q = \{q_1, q_2, \ldots, q_K\}$, $K \geq 1$, and Q is a subset of P.
- R: a set of L requested resource properties with their weights, $R = \{(r_1, w_1), (r_2, w_2), \ldots, (r_L, w_L)\}$, $L \geq 1$.

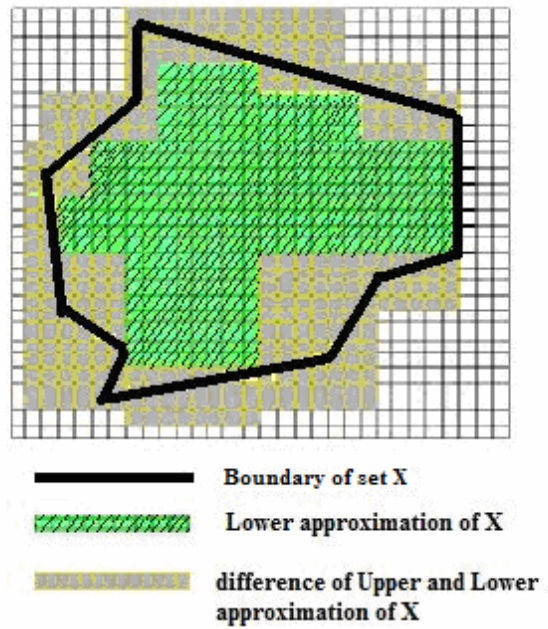

Figure 1. Approximation in rough set theory

According to the rough set theory for a given set X there are:

$$\underline{Q}X = \{x | [X]_Q \subseteq X\} \quad (1)$$

$$\overline{Q}X = \{x | [X]_Q \cap X \neq \phi\} \quad (2)$$

In which $\underline{Q}X$ is the lower approximation and $\overline{Q}X$ is the upper approximation of X in terms of properties set Q. X is a subset of U and Q is a sub set of P.

$$X \subseteq U$$
$$Q \subseteq P$$

So for a property $q \in Q$, we can say that:

- $\forall x \in \underline{Q}X$, x definitely is a member of X and definitely has property q.
- $\forall x \in \overline{Q}X$, x probably is member of X and probably has property q.

$\forall x \in U - \overline{X}$, x absolutely is not a member of X and absolutely does not have property q.

The Most important part of rough set theory is attribute reduction. Some attributes are dependent on other attributes in attributes set, so they are not necessary to be considered in matching phase. According to rough set theory we are:

$$POS_C(D) = \bigcup_{X \in U/D} \underline{C}X \quad (3)$$

$$\alpha = \gamma(C,D) = \frac{|\underline{C}X|}{|U|} \quad (4)$$

In Which C and D are subsets of property set P. as shown in [13], D totally depends on C if $\alpha = 1$ Or D partially (in a degree of $\alpha$) depends on C if $\alpha < 1$.

Since existing works need to find exact match between requested resources and registered resources, it is difficult to find exact matching. So by using rough set theory, the need of exact match has been removed.

## IV. DYNAMIC ROUGH SET THEORY

Although rough set theory is being used in various ranges of research such as data mining, pattern recognition, decision making and expert system, it is suitable for static knowledge and data. In fact, in a classical rough set theory, subset X of universal set U is a static set without considering the dynamic features it can have. In the real word, most information systems have dynamic features so that the rate of participant and disappearance of entities in these systems is high. Whereas Pawlak's rough set theory can only deal with static information system, using a dynamic method to deal with uncertainty and process information system will have more efficiency.

By using dynamic rough set theory, considering dynamic features of an information system will be possible. Dynamic rough set theory uses outward and inward transfer parameters to expand or contract X set in classical rough set.

According to [20], dynamic rough set theory has been defined as follows:

Suppose A= (U, P) is an information system, $T \subseteq P$ and $X \subseteq U$. For any $x \in U$, we have:

$$\rho^-_{(X,T)}(x) = \frac{|[x]_T - X|}{|[x]_T|}, \text{ as } x \in X \quad (5)$$

$$\rho^+_{(X,T)}(x) = 1 - \frac{|[x]_T - X|}{|[x]_T|}, \text{ as } x \in \sim X \quad (6)$$

$\rho^-_{(X,T)}(x)$ is called outward transfer coefficient and $\rho^+_{(X,T)}(x)$ is called inward transfer coefficient of element x about T. In real computation, outward and inward transfer coefficients are been choose as constant amounts. In fact $d^-_T(X) \in [0,1]$ and $d^+_T(X) \in [0,1]$ are outward transfer standard and inward transfer standard of elements of X about T, respectively.

Inflated dynamic main set of X is defined as below:

$$M^+_T(X) = \{x | x \in (\sim X), d^+_T(X) \leq \rho^+_{(X,T)} < 1\}. \quad (7)$$

And inflated dynamic assistant set is defined as:

$$A^+_T(X) = \{x | x \in (\sim X), 0 \leq \rho^+_{(X,T)} < d^+_T(X)\}. \quad (8)$$

$X^+_T$ is called inflated dynamic set of X about T and defined as:

$$X^+_T = X \cup M^+_T(X). \quad (9)$$

The formulas (5-9) show that we can expand X according to T. we can also contract X according to T. for this reason we have:

$$M^-_T(X) = \{x | x \in X, d^-_T(X) \leq \rho^-_{(X,T)}(X) < 1\}. \quad (10)$$

In which $M^-_T(X)$ is defined as contracted dynamic set of X about T and also contracted dynamic assistant set is defined as:

$$A^-_T(X) = \{x | x \in X, 0 \leq \rho^-_{(X,T)}(X) < d^-_T(X)\}. \quad (11)$$

And $X^-_T$ called contracted dynamic set is defined as:

$$X^-_T = X - M^-_T. \quad (12)$$

According to the above mentioned, we can expand and contract X according to T. Suppose we have $T$ and $T' \subseteq P$, two direction dynamic set of X according to the $T$ and $T'$ is defined:

$$X^*_{(T,T')} = (X - M^-_T(X)) \cup M^+_T(X). \quad (13)$$

Suppose $Q \subseteq P$, we can compute upper and lower approximation of $X^*_{(T,T')}$ using equations (1, 2) so that we have:

$$\underline{Q}^*_{(T,T')}(X) = \{x | x \in U, [x]_Q \subseteq X^*_{(T,T')}\}. \quad (14)$$

$$\overline{Q}^*_{(T,T')}(X) = \{x | x \in U, [x]_Q \cap X^*_{(T,T')}\} \quad (15)$$

$\underline{Q}^*_{(T,T')}(X)$ and $\overline{Q}^*_{(T,T')}(X)$ are called two direction transfer D-lower approximation set and two direction transfer D-upper approximation set of X, respectively.

In fact according to $M^+_T(X)$ we should increase resources (X) which can have opportunity of selection according to the attributes set T, but $M^-_{T'}(X)$ indicates according to the attributes set $T'$ we should decrease X.

$\underline{Q^*_{(T,T')}}(X)$ indicates the objects of the optimization of the candidate set which can be considered as a candidate set for matchmaking process. So in the matchmaking phase we only need to search D-lower approximation set ($\underline{Q^*_{(T,T')}}(X)$) in order to select resources which satisfy requested service.

In this work, we can also determine the priority of each requested service property, so that if properties T have an important role, their priority factor is high, we can decrease $d_T^+$ and this means that we expand candidate set X according to the properties set T. when $T'$ plays a less important role, priority of properties is low, we can decrease $d_{T'}^-$ in order to contract the candidate set.

## V. RESOURCE DISCOVERY

GridSim simulator has been used in order to simulate Dynamic Rough Set Resource Discovery Algorithm (DRSRD). As shown in Fig. 2, user sends a service request to the GridSim's Broker, Broker forwards the request to the GIS which can access Advertised Resource Repository and Ontology template in order to get resources which satisfy requested service. GIS has two components in order to find resources satisfying requested service. First component is Candidates Optimization which uses dynamic rough set theory in order to determine the optimum set of candidate resources. User defines a priority factor called Wi for each of the requested service properties in order to determine their priority. Candidate optimization component determines candidate resources set according to the priority of requested service properties.

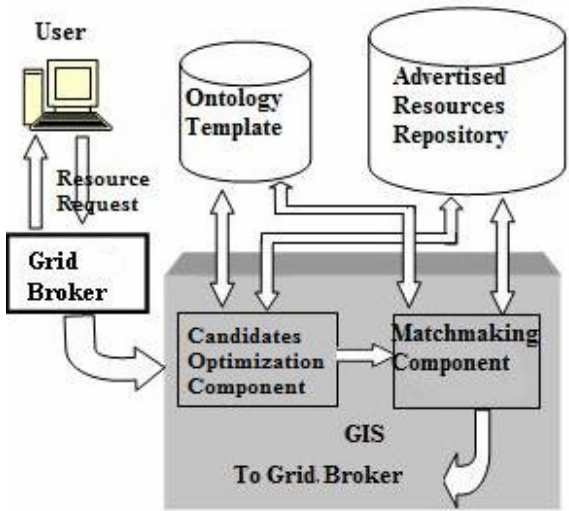

Figure 2. Algorithm outline

The Second component is the Matchmaking component which does the matchmaking algorithm on the candidate resources set obtained from the candidates optimization component.

For describing resource properties, we have used a resource ontology template based on the Karlsruhe ontology model [10]. The resource ontology template, as shown in Fig. 3 [23], has been created by considering the most possible computing resources in the Grid. The concept of these resources has been defined properly using relations and properties so that the characteristics of any resource can be defined by their properties. For using the ontology template in the GridSim, which is a java base simulator, we have used the protégé-OWL API, which is a java base API, in order to create and modify Ontology dynamically.

In this section we will describe the candidate optimization component and matchmaking component.

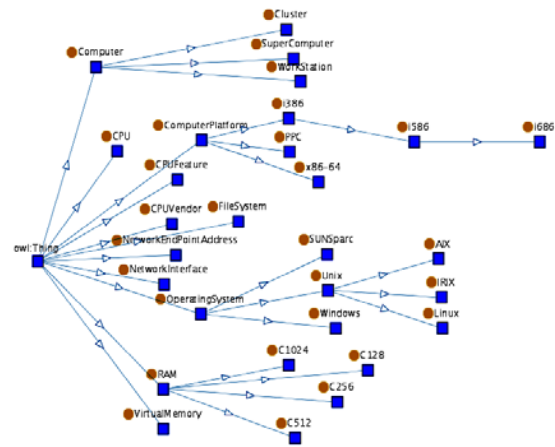

Figure 3. Ontology template

According to the method proposed in [14] there are four relations between pR and pA, in which pR and pA are respectively a property for the resource request and a property for the registered resource. These four relations are as follow:

- **Exact match,** $p_R$ and $p_A$ are equivalent or $p_R$ is a subclass of $p_A$.
- **Plug in match,** $p_A$ subsumes $p_R$.
- **Subsume match,** $p_R$ subsumes $p_A$.
- **No match,** no subsumption between $p_R$ and $p_A$.

Each property in the advertised resources properties set will be compared to all of the properties in the resource request properties set. Each property in the advertised resources that has no match relationship with any of properties in the resource request will be treated as an irrelevant property and must be marked up. This step must be repeated until all the properties in the registered resources have been checked. The marked up properties should not be used in the Candidates Optimization Component.

After the reduction of irrelevant properties, the remained properties will be sent to the Candidates Optimization Component entity to optimize candidates set.

### A. Candidates Optimization

The Most important aim of dynamic rough set theory is to deal with the vagueness and uncertainty in a knowledge system which changes dynamically. For a system such as the Grid whose resources can join or leave the system randomly, using dynamic rough set theory is more efficient than classical rough set theory.

User sends its service request to the Broker. In this request, each one of the requested service properties has a weight Wi. Broker forwards this request to the Grid Information Service (GIS) in order to find the best resources which satisfy the requested service. After getting the request by GIS, it classifies the requested properties according to their weight. According to part III, the set R is the requested resource properties and the properties set T, which $T \subseteq R$, is defined as bellow:

$$T = \{(r_i, w_i) | (r_i, w_i) \in R \text{ and } w_i \geq 0.5\}, 1 \leq i \leq L1.$$

In fact the set T contains properties with priority factor (weight) more than 0.5.

As mentioned in part IV the candidate set can be expanded according to the properties set T. According to the weight of requested service properties, we define the inward transfer standard $d_T^+(X)$ as follows:

$$d_T^+(X) = \frac{\sum_{i=1}^{L1} w_i}{|T|}, \text{which } (t_i, w_i) \in T \quad (16).$$

The properties set $T'$, in which $T' \subseteq R$, are defined as a set of properties the weight of which is less than 0.5. So $T'$ is defined as:

$$T' = \{(r_i, w_i) | (r_i, w_i) \in R \text{ and } w_i < 0.5\}, 1 \leq i \leq L2.$$

The outward transfer standard $d_{T'}^-(X)$ is defined as bellow:

$$d_{T'}^-(X) = \frac{\sum_{i=1}^{L2} w_i}{|T'|}, \text{which } (t_i, w_i) \in T' \quad (17).$$

The candidates set X is defined as a set of resources with maximum non empty properties according to the requested resource properties. And ~X is defined as all resources in the universal set U which are not contained in the X.

Candidates Optimization algorithm is shown in the Fig. 3. Algorithm uses three steps to compute candidates optimized set.

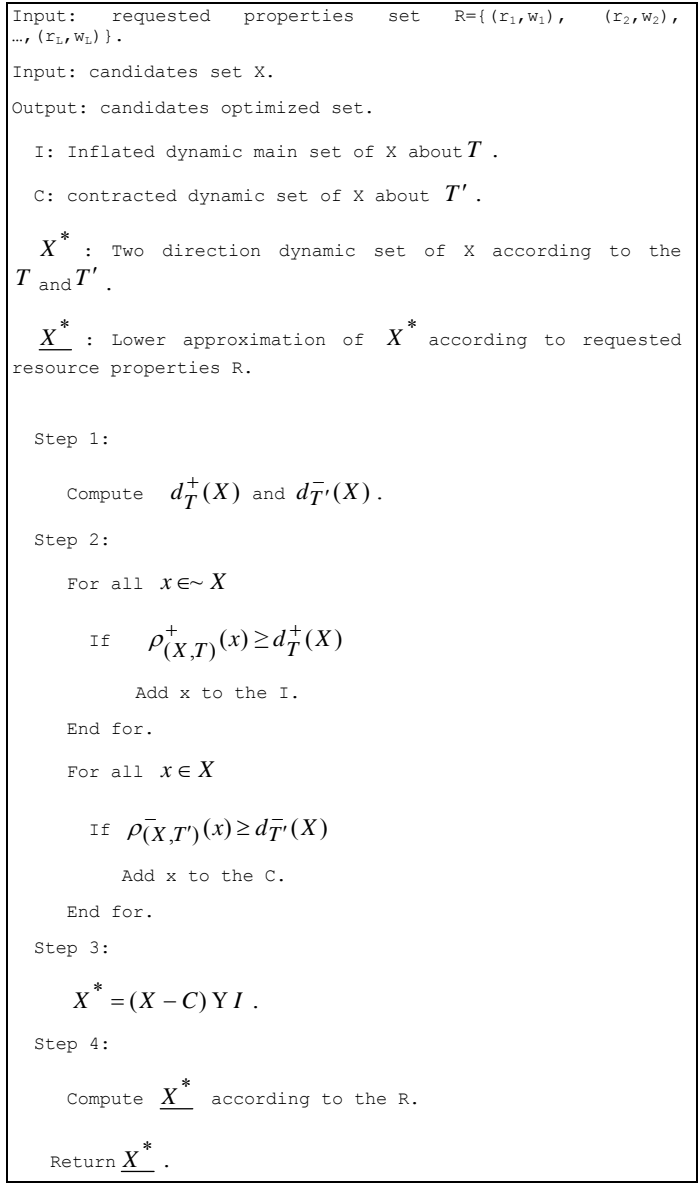

```
Input: requested properties set R={(r₁,w₁), (r₂,w₂),
..., (r_L,w_L)}.
Input: candidates set X.
Output: candidates optimized set.
  I: Inflated dynamic main set of X about T .
  C: contracted dynamic set of X about T' .
  X* : Two direction dynamic set of X according to the
T and T' .
  X*_ : Lower approximation of X* according to requested
resource properties R.

Step 1:
    Compute d_T^+(X) and d_{T'}^-(X) .
Step 2:
    For all x ∈~ X
      If  ρ_{(X,T)}^+(x) ≥ d_T^+(X)
        Add x to the I.
    End for.
    For all x ∈ X
      If  ρ_{(X,T')}^-(x) ≥ d_{T'}^-(X)
        Add x to the C.
    End for.
Step 3:
    X* = (X − C) Y I .
Step 4:
    Compute X*_ according to the R.
  Return X*_ .
```

Figure 4. Candidates Optimization algorithm

Step 1 calculates $d_T^+(X)$ and $d_{T'}^-(X)$ using the equations (16) and (17) respectively. In step 2, the inflated dynamic main set of X and contracted dynamic main set of X using equations (7) and (10) respectively.

Step 3 calculates two direction dynamic set of X according to $T$ and $T'$ using equation (13). Candidates set X can be expanded according to the properties set T which has properties with higher priority and can be contracted according to the properties set $T'$ the properties of which have lower priority. In Step 4, by using equation (14), the lower approximation set $\underline{x}^*$ of $X^*$ is calculated according to the requested resource properties set R. In fact, $\underline{x}^*$ is the set of resources that are most likely to be selected as the matched resources.

### B. Resource Matchmaking

After optimization of the candidates set we should only apply the matchmaking algorithm on the optimized candidates set. Reduction of the candidates set causes the reduction of searching time.

We design the matching resource algorithm according to the rules proposed in [14] and in regarding to the ontology template.

We define m(r$_i$, q$_j$) as the match degree of the requested resource property r$_i$ and the advertised resource property q$_j$. In this algorithm properties are divided in the two classes. The first class is properties with String type. For this class of properties if q$_j$ is an exact match with r$_i$ the match degree is 1.0. But if q$_j$ is a plug in match with r$_i$ with a match generation of d:

$$\begin{cases} m(q_j, r_i) = 1 - ((d-1) \times 0.1) & 2 \leq d \leq 5 \\ m(q_j, r_i) = 0.5 & d > 5 \end{cases}$$

For the case of the subsume match if q$_j$ is a subsume match with r$_i$ with the match generation of d:

$$\begin{cases} m(q_j, r_i) = 0.8 - ((d-1) \times 0.1) & 1 \leq d \leq 3 \\ m(q_j, r_i) = 0.5 & d > 3 \end{cases}$$

An advertised property with empty value is regarded as null property. For any null property q$_j$ the match degree is 0.5.

The second class is properties set with Non string type. This class contains properties with type integer, long integer and double. For this class if type of both properties is equal, match degree is defined by:

$$\begin{cases} m(q_j, r_i) = 1 - (V(q_j)/V(r_i) \times 0.1) & V(q_j)/V(r_i) \leq 5 \\ m(q_j, r_i) = 0.5 & V(q_j)/V(r_i) > 5 \end{cases}$$

In which V(q$_j$) is the value of attribute q$_j$.

Fig. 5 shows conditions for calculating the match degree.

For calculating the match between the requested resource and the advertised resource we have used the equation (18) which calculates the maximum match degree between the requested resource and the advertised resource.

$$M(R_R, R_A) = \sum_{i=1}^{L} \left( \sum_{j=1}^{K} MAX(m(q_j, r_i)) \times w_i \right) \bigg/ \sum_{i=1}^{L} w_i \quad (18).$$

In the formula (18), the symbols R$_R$ and R$_A$ are the requested resource and the advertised resource respectively.

According to this algorithm, matched resources can be ranked according to their match degree. Ranking process is done according to the priority of properties.

```
For each  q_j ∈ Q  &&  V(q_j) ≠ null
  For each  r_i ∈ R
    If type of both  q_j && r_i is string
      If  q_j is an exact match with  r_i
        m(q_j, r_i) = 1.0.
      Else if  q_j is an plug in match with  r_i and match
degree d
        If 2 ≤ d ≤ 5
          m(q_j, r_i) = 1 - (d - 1) × 0.1.
        Else if d>5
          m(q_j, r_i) = 0.5.
      Else if  q_j is an subsume match with  r_i
        If  q_j is  d_th subclass of  r_i
          If 1≤d≤3
            m(q_j, r_i) = 0.8 - (d - 1) × 0.1.
          Else if d >3
            m(q_j, r_i) = 0.5.
    Else if type of both  q_j && r_i is not string and is
equal
      If  V(q_j)/V(r_i) ≤ 5
        m(q_j, r_i) = 1 - (V(q_j)/ V(r_i) * 0.1) .
      Else
        m(q_j, r_i) = 0.5.
  End for
End for
For each  q_j ∈ Q  &&  V(q_j) = null
  For each  q_i ∈ Q
    m(q_j, r_i) = 0.5.
  End for
End for
```

Figure 5. match degree algorithm

## VI. EXPERIMENTAL RESULTS

In order to simulate algorithm we run the GridSim that is a grid java based simulator. We have also used db4o [22] data base as a repository for advertised resources. We have created ontology of possible resources using protégé API [10], which is a java based API, for semantic description of resources. The structure of the ontology of resources is motivated by the need to provide information about resources. The resource ontology proposed in this paper takes most of computing resources into account. This ontology template has been created according to the basis of Karlsruhe Ontology model [23].

In order to test our algorithm we simulated 10000 resources which are semantically defined according to the ontology template shown in Fig. 3. Each resource register itself at the database as soon as joined the grid by sending its features which are defined according to the ontology template. For designing Query generator we created users which send resource requests with deferent requested resource property. Requested resource properties are defined according to the ontology template.

As shown in Fig. 6, user sends its resource request to the GridSim's broker. Broker forwards this resource request to the Grid Information Server (GIS). The GIS uses ontology and accesses the database in order to find advertised resources relevant to the requested resource. Retrieved resources ID along with its match degree are sent back to the user.

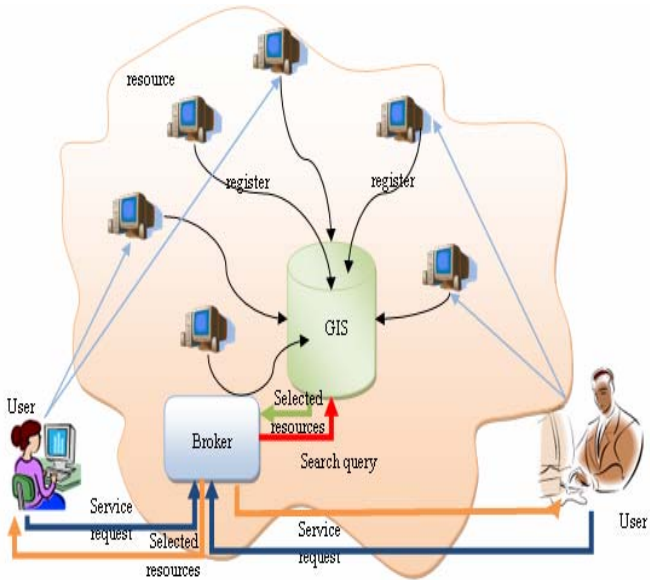

Figure 6. GridSim Architecture

We have tested our algorithm with resource property certainty of 30%, 50%, 80%, and 100% and for each of these states we have run simulator with deferent number of advertised resources. We have used average results of 100 times run of each case for comparison.

For evaluating precision and matching time of our algorithm we compared this algorithm with the algorithm proposed in [14] which is a combination of UDDI and OWL-S and rough set based algorithm proposed in our previous work [18].

### A. Precision evaluation

As mentioned above we test our algorithm with 4 groups of advertised resources. The First group has only 30% properties certainty. The Second group has 50% property certainty and the third group has 80% property certainty and the fourth group has 100% property certainty. Fig. 7 to Fig. 13 show the comparison of the precision for different numbers of the resources. Precision is defined as the ratio of the number of correct retrieved resources rather than all the retrieved resources. According to matching algorithm proposed in [14], UDDI and OWL-S matching algorithm can not deal with uncertainty.

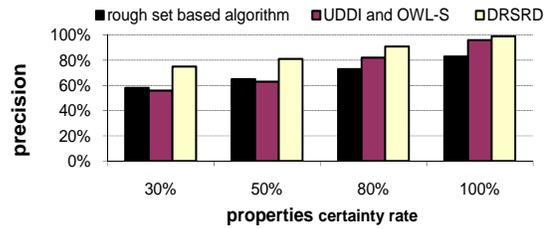

Figure 7. comparison of precision for 500 resources

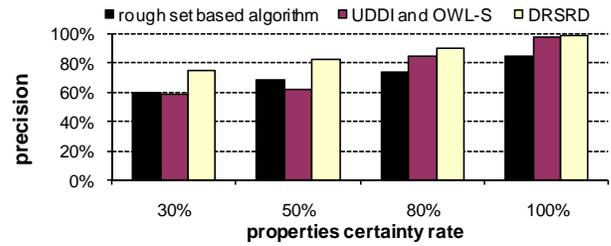

Figure 8. comparison of precision for 1000 resources

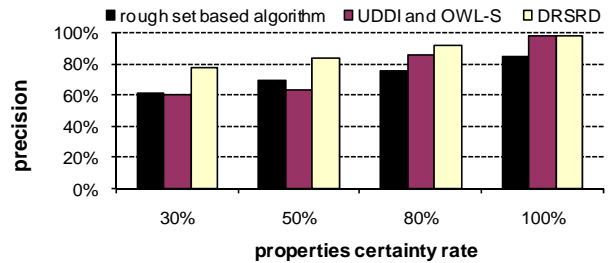

Figure 9. comparison of precision for 2000 resources

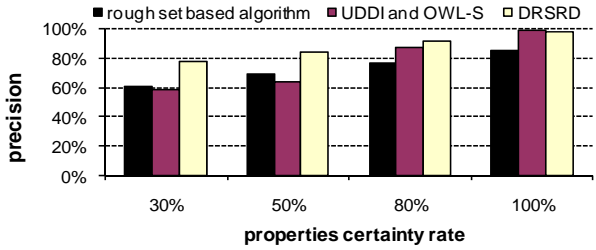

Figure 10. comparison of precision for 4000 resources

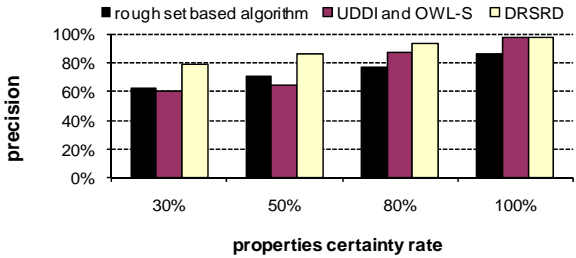

Figure 11. comparison of precision for 6000 resources

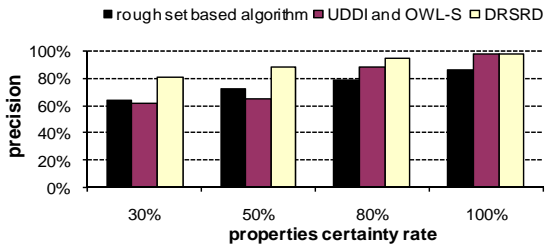

Figure 12. comparison of precision for 8000 resources

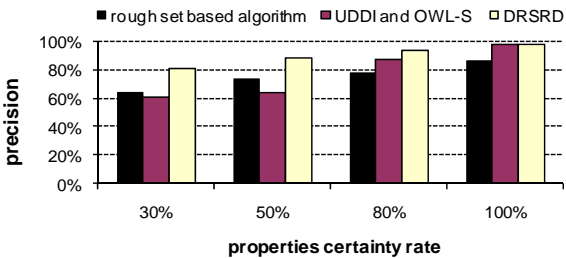

Figure 13. comparison of precision for 10000 resources

As shown in the above figures, the precision of the combination of UDDI and OWL-S is lower than Dynamic Rough Set Resource discovery (DRSRD) algorithm for 30%, 50%, and 80% of service property certainty. This is because of disability of UDDI and OWL-S in dealing with uncertainty.

Also the precision of DRSRD is more than rough set based algorithm. This is because of the dynamic features of the Grid environment. Whereas classic rough set theory can not deal with dynamic features, rough set based algorithm has low precision. By increasing the certainty, deference between UDDI and OWL-S combined algorithm and DRSRD algorithm is being decreased so that with 100% certainty the precision of both of two algorithms reaches 100%. But for different rates of certainty, DRSRD is more precise than rough set based algorithm. It is clear that DRSRD has a good effect on dealing with vagueness and dynamic features of grid.

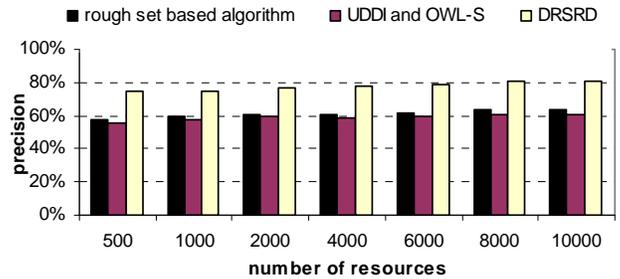

Figure 14. Precision increment for 30% certainty rate

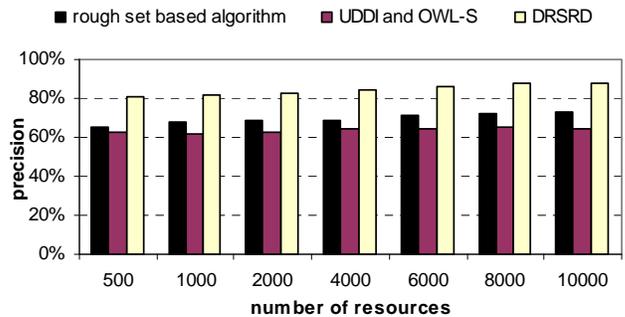

Figure 15. Precision increment for 50% certainty rate

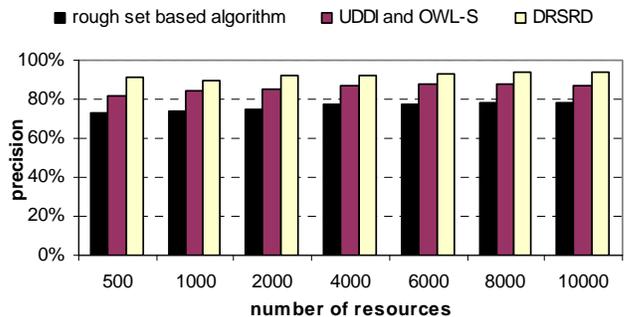

Figure 16. Precision increment for 80% certainty rate

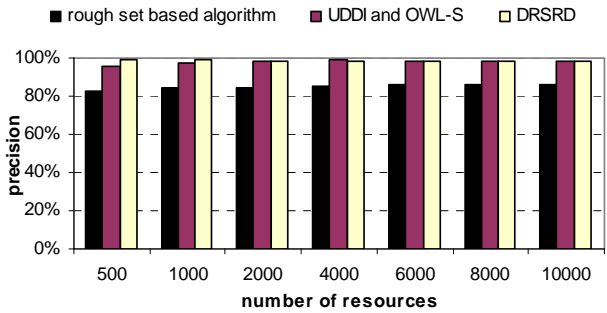

Figure 17. Precision increment for 100% certainty rate

Fig. 14 to Fig. 17 show the increment of precision according to the increment of the number of the resources for 30%, 50%, 80%, and 100% certainty rate, respectively. Along with the increase of the number of resources, precision also increases. It is because of the increasing of the survey population. Define abbreviations and acronyms the first time they are used in the text, even after they have been defined in the abstract. Abbreviations such as IEEE, SI, MKS, CGS, sc, dc, and rms do not have to be defined. Do not use abbreviations in the title or heads unless they are unavoidable.

*B. Matching time evaluation*

For evaluating matching time we run our simulator 100 times with different amount of advertised resources. We have compared DRSRD algorithm with rough set based algorithm and UDDI and OWL-S combined model to evaluate the matching time of our algorithm.

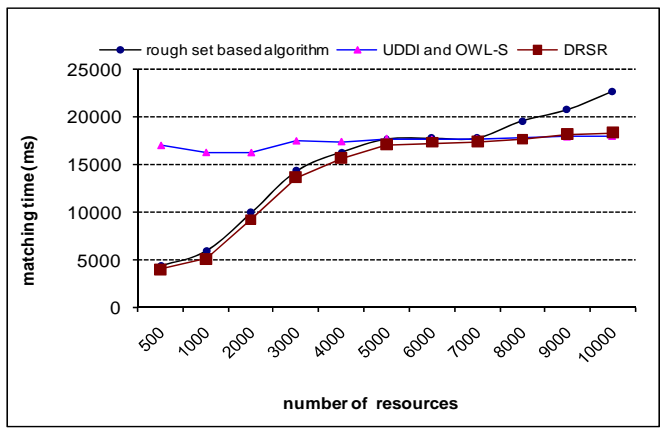

Figure 18. Comparison of the matching time

Fig. 18 shows that the matching time of DRSRD algorithm is lower than UDDI and OWL-S when the number of advertised resources is under 9000. By increasing the number of advertised resources UDDI and OWL-S combined model is more effective because its matching time depends on number of properties rather than number of advertised resources. It is also clear that DRSRD has lower matching time rather than rough set based algorithm.

## VII. CONCLUSION AND FUTURE WORK

In this paper we have shown dynamic rough set based algorithm has a good effect in dealing with uncertainty and vagueness for resource matching in a dynamic environment such as grid. Using classic rough set theory in order to deal with vagueness is effective but it is only for static data. Whereas grid is a dynamic environment and features of resources change dynamically, we need to use a dynamic method to deal with vagueness, so we have used dynamic rough set theory. DRSRD algorithm can deal with uncertain properties and find a set of resource which may maximally satisfy the needs of requested resource. In fact, our algorithm can find a list of resources which have high degree of matching according to the weight of requested properties. Experimental results have shown that the DRSRD algorithm is more effective in resource matching than rough set based algorithm and UDDI and OWL-S combined algorithm. Algorithm time for our algorithm is lower than rough set based algorithm. It is also lower than UDDI and OWL-S algorithm for resources number less than 9000 resources.